\documentstyle[preprint,aps]{revtex}
\begin{document}
\draft
\title{Dilaton Black Holes in de Sitter or Anti-de Sitter Universe}
\author{Chang Jun Gao$^{1}$\thanks{E-mail: gaocj@mail.tsinghua.edu.cn}
Shuang Nan Zhang$^{1,2,3,4}$\thanks{E-mail:
zhangsn@mail.tsinghua.edu.cn}}
\address{$^{1}$Department of Physics and Center for Astrophysics, Tsinghua University, Beijing 100084, China(mailaddress)}
\address{$^2$Physics Department, University of Alabama in Huntsville, AL 35899, USA }
\address{$^3$Space Science Laboratory, NASA Marshall Space Flight Center, SD50, Huntsville, AL 35812, USA }
\address{$^4$Laboratory for Particle Astrophysics, Institute of High Energy Physics, Chinese Academy of Sciences, Beijing 100039, China}

\date{\today}
\maketitle

\begin{abstract}
\hspace*{7.5mm}Poletti and Wiltshire have shown that, with the
exception of a pure cosmological constant, the solution of a
dilaton black hole in the background of de Sitter or anti-de
Sitter universe, does not exist in the presence of one
Liouville-type dilaton potential. Here with the combination of
three Liouville-type dilaton potentials, we obtain the dilaton
black hole solutions in the background of de Sitter or anti-de
Sitter universe.
\end{abstract}

\section{introduction}
 \hspace*{7.5mm}Dilaton is a kind of scalar field occurring in the
 low energy limit of the string theory where the Einstein action is supplemented
 by fields such as axion, gauge fields and dilaton coupling in a
 nontrivial way to the other fields. Exact solutions for charged dilaton black holes in which
 the dilaton is coupled to the Maxwell field have been constructed
 by many authors. It is found that the presence of dilaton has
 important consequences on the causal structure and the
 thermodynamic properties of the black hole [1-9]. Thus much
 interest has been focused on the study of the dilaton black
 holes.\\
 \hspace*{7.5mm}On the other hand, there has also been some
 renewed interest in the study of black hole theories with the
 cosmological constant. Theories with negative cosmological constant
 can be embedded in a supersymmetric setting in which gauged supergravity theories are obtained in
various dimensions. Gauged supergravities admit anti-de Sitter
spacetime as a vacuum state, and thus black hole solutions of
these theories are of physical relevance to the proposed AdS/CFT
correspondence [10-15]. In particular, the study of AdS black
holes can give new insights into the nonperturbative structure of
some conformal field theories. Black holes in the background with
positive cosmological constant, i.e., in de Sitter spacetime, have
also attracted some interest recently, due to the phenomenon of
black hole anti-evaporation [16]. Besides, these black holes could
be relevant to the proposed duality between the large N limit of
Euclidean four-dimensional U(N) super-Yang-Mills theory and the
so-called type IIB. string theory in de Sitter spacetime [17]. So
the object of the present paper is to find the dilaton black hole
solutions in the de Sitter or anti-de Sitter spacetime.\\
\hspace*{7.5mm}In fact, Poletti and Wiltshire [18] have shown that
with the exception of a pure cosmological constant, no dilaton-de
Sitter or anti-de Sitter black hole solution exists with the
presence of only one Liouville-type dilaton potential. Okai, who
made investigations using power series [19], has been unable to
prove unequivocally that dilaton-de Sitter or anti-de Sitter black
hole solutions do exist. In this paper, with the combination of
three Liouville-type dilaton potentials, the solutions of dilaton
black holes in the background
of de Sitter or anti-de Sitter spacetime are achieved.\\
\hspace*{7.5mm}The paper is organized as follows. In Sec.2, we
will derive the solution in the cosmic coordinate system with a
new method developed recently by us. In Sec.3, we find the
coordinates transformation which recast the solution in the
schwarzschild coordinates system. In Sec.4, we deduce the dilaton
potential with respect to the cosmological constant. In Sec.5 and
Sec.6, we generalize the solution to arbitrary coupling constant
and arbitrary number black holes cases. In Sec.7, the horizon
property of the black hole is discussed. We conclude with some
final remarks.
\section{dilaton-de Sitter metric in cosmic
coordinate system} \hspace*{7.5mm}The metric of a dilaton black
hole in the Schwarzschild coordinate system is given by
\begin{equation}
d{s}^2=-\left(1-\frac{2M}{{x}}\right)
d{t}^2+\left(1-\frac{2M}{{x}}\right)^{-1}d{x}^2+x\left(x-2D\right)d\Omega_{2}^2,
\end{equation}
where $M$ and $D$ are the mass and dilaton charge of the black
hole, respectively. $D$ is related to the mass $M$ and electric
charge $Q$ as follows
\begin{equation}
D=\frac{Q^2e^{2\phi_0}}{2M}.
\end{equation}
$\phi_0$ is the asymptotic constant value of the dilaton.\\
\hspace*{7.5mm}In order to obtain the dilaton-de Sitter metric, we
should rewrite the metric Eq.(1) in the cosmic coordinate system.
So make variable transformation $x\rightarrow r$
 \begin{equation}
{x}=\frac{\left(r+M+D\right)^2-4MD}{2r},
\end{equation}
 then we can rewrite Eq.(1) as follows
\begin{eqnarray}
ds^2&=&-\frac{\left(1-\frac{M}{r}+\frac{D}{r}\right)^2}{\left(1+\frac{M}{r}+\frac{D}{r}\right)^2
-\frac{4MD}{r^2}}dt^2\nonumber\\&&
+\frac{1}{4}{\left(1+\frac{M}{r}-\frac{D}{r}\right)^2}\left[\left(1+\frac{M}{r}+\frac{D}{r}\right)^2
-\frac{4MD}{r^2}\right]\left(dr^2+r^2d\Omega_{2}^2\right).
\end{eqnarray}
Re-scale the variables $t$ and $s$, we have
\begin{eqnarray}
ds^2&=&-\frac{\left(1-\frac{M}{r}+\frac{D}{r}\right)^2}{\left(1+\frac{M}{r}+\frac{D}{r}\right)^2
-\frac{4MD}{r^2}}dt^2\nonumber\\&&
+{\left(1+\frac{M}{r}-\frac{D}{r}\right)^2}\left[\left(1+\frac{M}{r}+\frac{D}{r}\right)^2
-\frac{4MD}{r^2}\right]\left(dr^2+r^2d\Omega_{2}^2\right).
\end{eqnarray}
\hspace*{7.5mm}Following the method we developed recently [20], we
make the following replacements
\begin{eqnarray}
1\longrightarrow {{a^{\frac{1}{2}}}},\ \ \ \
\frac{M}{r}\longrightarrow \frac{M}{ra^{\frac{1}{2}}},\ \ \ \
\frac{D}{r}\longrightarrow \frac{D}{ra^{\frac{1}{2}}},
\end{eqnarray}
where $a\equiv e^{Ht}$, $H$ is a constant which has the meaning of
Hubble constant. Then the dilaton-de Sitter metric is achieved
\begin{eqnarray}
ds^2&=&-\frac{\left(1-\frac{M}{ar}+\frac{D}{ar}\right)^2}{\left(1+\frac{M}{ar}+\frac{D}{ar}\right)^2
-\frac{4MD}{a^2r^2}}dt^2\nonumber\\&&
+a^2{\left(1+\frac{M}{ar}-\frac{D}{ar}\right)^2}\left[\left(1+\frac{M}{ar}+\frac{D}{ar}\right)^2
-\frac{4MD}{a^2r^2}\right]\left(dr^2+r^2d\Omega_{2}^2\right).
\end{eqnarray}
Eq.(7) is just the charged dilaton black hole solution in the
background of de Sitter universe in the cosmic coordinate system.
When $M=D=0$, it recovers the well-known de Sitter metric. On the
other hand, when $H=0$, it recovers the charged dilaton metric
Eq.(5). In the next section, we rewrite it in the Schwarzschild
coordinate system by coordinates transformation.
\section{dilaton-de Sitter or anti-de Sitter metrics in Schwarzschild coordinate system}
\hspace*{7.5mm}In order to present the dilaton-de Sitter metric in
the Schwarzschild coordinate system, we make variable
transformation $r\rightarrow y$ as follows
\begin{equation}
r=a^{-1}\left[y-M-D+\sqrt{\left(y-2M\right)\left(y-2D\right)}\right].
\end{equation}
Eq.(7) becomes
\begin{eqnarray}
d{s}^2&=&-\left[1-\frac{2M}{{y}}-4y\left(y-2D\right)H^2\right]
d{t}^2+\frac{4}{1-\frac{2M}{{y}}}d{y}^2\nonumber\\&&-\frac{8H\sqrt{y\left(y-2D\right)}}
{\sqrt{1-\frac{2M}{y}}}dtdy+4y\left(y-2D\right)d\Omega_{2}^2.
\end{eqnarray}
Re-scale $t$ and $s$, we have
\begin{eqnarray}
d{s}^2&=&-\left[1-\frac{2M}{{y}}-4y\left(y-2D\right)H^2\right]
d{t}^2+\frac{1}{1-\frac{2M}{{y}}}d{y}^2\nonumber\\&&-\frac{4H\sqrt{y\left(y-2D\right)}}
{\sqrt{1-\frac{2M}{y}}}dtdy+y\left(y-2D\right)d\Omega_{2}^2.
\end{eqnarray}
Eq.(10) has a $dtdy$ term. In order to eliminate this term, we
introduce a new time variable $u$, namely, $t\rightarrow u$
\begin{equation}
t=u-\int{\frac{2H\sqrt{y\left(y-2D\right)}}{\sqrt{1-\frac{2M}{y}}\left[1-\frac{2M}{{y}}-4y\left(y-2D\right)H^2\right]
}dy}.
\end{equation}
Then Eq.(10) is reduced to
\begin{eqnarray}
d{s}^2&=&-\left[1-\frac{2M}{{y}}-4y\left(y-2D\right)H^2\right]
d{u}^2+\left[1-\frac{2M}{{y}}-4y\left(y-2D\right)H^2\right]^{-1}d{y}^2\nonumber\\&&+y\left(y-2D\right)d\Omega_{2}^2.
\end{eqnarray}
Let $H^2$ absorb the constant $4$ and rewrite the variables
$(t,r)$ instead of $(u,y)$, we obtain the dilaton-de Sitter metric
in the Schwarzschild coordinate system
\begin{eqnarray}
d{s}^2&=&-\left[1-\frac{2M}{{r}}-r\left(r-2D\right)H^2\right]
d{t}^2+\left[1-\frac{2M}{{r}}-r\left(r-2D\right)H^2\right]^{-1}d{r}^2\nonumber\\&&+r\left(r-2D\right)d\Omega_{2}^2.
\end{eqnarray}
When $D=0$, it restores to the well-known Schwarzschild-de Sitter
metric. On the other hand, when $H=0$, it restores to the charged
dilaton metric which is found by Garfinkle, Horowitz and
Strominger [2]. To show Eq.(13) does represent a solution of
Einstein-Maxwell-dilaton theory, in the following section, we will
derive the potential of the dilaton for the cosmological constant.
In order that the negative cosmological constant case is included
in our solution, we make the replacement of $H^2=\lambda/3$ in
Eq.(13). $\lambda$ is the cosmological constant
which can be positive or negative.\\
\section{Potential of the dilaton for cosmological constant}
 \hspace*{7.5mm}We consider the four-dimensional theory in which
 gravity is coupled to dilaton and Maxwell field with an action
\begin{eqnarray}
S=\int{d^4x\sqrt{-g}\left[R-2\partial_{\mu}\phi\partial^{\mu}\phi-V\left(\phi\right)-e^{-2\phi}F^2\right]},
\end{eqnarray}
where $R$ is the scalar curvature, $F^2=F_{\mu\nu}F^{\mu\nu}$ is
the usual Maxwell contribution, and $V\left(\phi\right)$ is a
potential for $\phi$.\\
\hspace*{7.5mm}Varying the action with respect to the metric,
Maxwell, and dilaton fields, respectively, yields
\begin{equation}
R_{\mu\nu}=2\partial_{\mu}\phi\partial_{\nu}\phi+\frac{1}{2}g_{\mu\nu}V+2e^{-2\phi}\left(F_{\mu\alpha}
F_{\nu}^{\alpha}-\frac{1}{4}g_{\mu\nu}F^2\right),
\end{equation}
\begin{equation}
\partial_{\mu}\left(\sqrt{-g}e^{-2\phi}F^{\mu\nu}\right)=0,
\end{equation}
\begin{equation}
\partial_{\mu}\partial^{\mu}\phi=\frac{1}{4}\frac{\partial V}{\partial \phi}-\frac{1}{2}e^{-2\phi}F^2.
\end{equation}
\hspace*{7.5mm}The most general form of the metric for the static
space-time can be written as
\begin{equation}
ds^2=-U\left(r\right)dt^2+\frac{1}{U\left(r\right)}dr^2+f\left(r\right)^2d\Omega_2^2.
\end{equation}
Then the Maxwell equation Eq.(16) can be integrated to give
\begin{equation}
F_{01}=\frac{Qe^{2\phi}}{f^2},
\end{equation}
where $Q$ is the electric charge. With the metric Eq.(18) and the
Maxwell field Eq.(19), the equations of motion Eqs.(15-17) reduce
to three independent equations
\begin{equation}
\frac{1}{f^2}\frac{d}{dr}\left(f^2U\frac{d\phi}{dr}\right)=\frac{1}{4}\frac{dV}{d\phi}+e^{2\phi}\frac{Q^2}{f^4},
\end{equation}
\begin{equation}
\frac{1}{f}\frac{d^2f}{dr^2}=-\left(\frac{d\phi}{dr}\right)^2,
\end{equation}
\begin{equation}
\frac{1}{f^2}\frac{d}{dr}\left(2Uf\frac{df}{dr}\right)=\frac{2}{f^2}-V-2e^{2\phi}\frac{Q^2}{f^4}.
\end{equation}
Substituted
\begin{equation}
f=\sqrt{r\left(r-2D\right)},\ \ \ \
U=1-\frac{2M}{r}-\frac{1}{3}\lambda r\left(r-2D\right),
\end{equation}
into Eqs.(20-22), we obtain the dilaton field, dilaton charge and
dilaton potential
\begin{eqnarray}
&&e^{2\phi}=e^{2\phi_0}\left(1-\frac{2D}{r}\right),\nonumber\\
&&D=\frac{Q^2e^{2\phi_0}}{2M},\nonumber\\ &&
V\left(\phi\right)=\frac{4}{3}\lambda+\frac{1}{3}\lambda
\left[e^{2\left(\phi-\phi_0\right)}+e^{-2\left(\phi-\phi_0\right)}\right].
\end{eqnarray}
Compare these solutions to the result of Garfinkle, Horowitz and
Strominger, we find that the Maxwell field, the dilaton filed and
the dilaton charge of the (anti) de Sitter versions are exactly
identical to that of the dilaton black
hole. The potential of the dilaton is the combination of a constant and two Liouville-type potentials. \\
\hspace*{7.5mm}So far, we obtained the action of the
dilaton-(anti) de Sitter black hole
\begin{eqnarray}
S=\int{d^4x\sqrt{-g}\left\{R-2\partial_{\mu}\phi\partial^{\mu}\phi-
\frac{4}{3}\lambda-\frac{1}{3}\lambda
\left[e^{2\left(\phi-\phi_0\right)}+e^{-2\left(\phi-\phi_0\right)}\right]-e^{-2\phi}F^2\right\}}.
\end{eqnarray}
When $\phi=\phi_0=0$, it reduces to the action of
Reissner-Nordstr$\ddot{o}$m-de Sitter black hole.
\section{dilaton-(anti) de Sitter metric with arbitrary $\alpha$}
\hspace*{7.5mm}Up to now, we have only dealt with the
dilaton-(anti) de Sitter metric for the case that the coupling
constant $\alpha=1$. In this section, we will present the
dilaton-(anti) de Sitter metric with an arbitrary coupling
constant. Following the method adopted above, we obtain the
dilaton-(anti) de Sitter metric with an arbitrary coupling
constant $\alpha$ in the cosmic coordinate system
\begin{eqnarray}
ds^2&=&-{\left(1-\frac{r_+}{ar}+\frac{r_{-}}{ar}\right)^2\left(1+\frac{r_+}{ar}-\frac{r_{-}}{ar}\right)^
{\frac{2\left(1-\alpha^2\right)}{1+\alpha^2}}}\left[{\left(1+\frac{r_+}{ar}+\frac{r_-}{ar}\right)^2
-\frac{4r_+r_-}{a^2r^2}}\right]^{\frac{-2}{1+\alpha^2}}dt^2\nonumber\\&&
+a^2\left(1+\frac{r_+}{ar}-\frac{r_{-}}{ar}\right)^{\frac{4\alpha^2}{1+\alpha^2}}\left[{\left(1+\frac{r_+}{ar}+\frac{r_-}{ar}\right)^2
-\frac{4r_+r_-}{a^2r^2}}\right]^{\frac{2}{1+\alpha^2}}\left(dr^2+r^2d\Omega_{2}^2\right),
\end{eqnarray}
where $a=e^{Ht}$ is the scale factor of the universe and the
equivalent form in Schwarzschild coordinate system
\begin{eqnarray}
d{s}^2&=&-\left[\left(1-\frac{r_{+}}{{r}}\right)\left(1-\frac{r_{-}}{{r}}\right)^{\frac{1-\alpha^2}{1+\alpha^2}}
-\frac{1}{3}\lambda
r^2\left(1-\frac{r_{-}}{r}\right)^{\frac{2\alpha^2}{1+\alpha^2}}\right]
d{t}^2\nonumber\\&&+\left[\left(1-\frac{r_{+}}{{r}}\right)\left(1-\frac{r_{-}}{{r}}\right)^{\frac{1-\alpha^2}{1+\alpha^2}}
-\frac{1}{3}\lambda
r^2\left(1-\frac{r_{-}}{r}\right)^{\frac{2\alpha^2}{1+\alpha^2}}\right]^{-1}d{r}^2\nonumber\\&&+
r^2\left(1-\frac{r_{-}}{r}\right)^{\frac{2\alpha^2}{1+\alpha^2}}d\Omega_{2}^2,
\end{eqnarray}
where $r_{+}$, $r_{-}$ are two event horizons of the black hole.
$\alpha$ is an arbitrary constant governing the strength of the
coupling between the dilaton and the Maxwell field. The action
with respect to the metric is
\begin{eqnarray}
S&=&\int
d^4x\sqrt{-g}\left\{R-2\partial_{\mu}\phi\partial^{\mu}\phi-e^{-2\alpha\phi}F^2
\right.\nonumber\\
&&\left.-\frac{2}{3}\lambda\frac{1}{\left(1+\alpha^2\right)^2}\left[\alpha^2\left(3\alpha^2-1\right)e^{-2\phi/\alpha}
+\left(3-\alpha^2\right)e^{2\phi\alpha}+8\alpha^2e^{\phi\alpha-\phi/\alpha}\right]\right\}.
\end{eqnarray}
When $\alpha=1$, the action restores to the
Garfinkle-Horowitz-Strominger case. The corresponding dilaton
field is
\begin{equation}
e^{2\alpha\phi}=e^{2\alpha\phi_0}\left(1-\frac{r_{-}}{r}\right)^{\frac{2\alpha^2}{1+\alpha^2}},\
\ \ \
e^{2\alpha\phi_0}=\frac{r_+r_{-}}{\left(1+\alpha^2\right)Q^2}.
\end{equation}
It is apparent that three Liuoville-type potentials constitute the
potential of the dilaton with respect to the cosmological
constant.

\section{multi-dilaton-black-hole solution in de Sitter universe}
\hspace*{7.5mm}In what follows we shall, for completeness, present
the multi-black hole solution in de Sitter universe. For extreme
dilaton black hole, set $r_+=r_-=m/4$, Eq.(26) becomes
\begin{eqnarray}
ds^2&=&-\left(1+\frac{m}{ar}\right) ^{\frac{-2}{1+\alpha^2}}dt^2
+a^2\left(1+\frac{m}{ar}\right)
^{\frac{2}{1+\alpha^2}}\left(dr^2+r^2d\Omega_{2}^2\right),
\end{eqnarray}
It closely resembles the Kastor-Traschen [21] solution and
Gibbons-Kallosh [22] solution. Thus following the method developed
by Kastor and Traschen, we obtain the cosmological
multi-dilaton-black hole solution
\begin{eqnarray}
ds^2&=&-\Omega^{\frac{-2}{1+\alpha^2}}dt^2+a\left(t\right)^2\Omega^{\frac{2}{1+\alpha^2}}\left(dx^2+dy^2+dz^2\right),
\nonumber\\ \Omega &=&1+\sum_i\frac{m_{i}}{ar_{i}}, \ \ \ \
r_i=\sqrt{\left(x-x_i\right)^2+\left(y-y_i\right)^2+\left(z-z_i\right)^2},\
\ \ \ a\left(t\right)=e^{Ht},
\end{eqnarray}
where $x, y, z$ are Cartesian coordinates on $R^3$, $m_i$ is the
mass or charge of the $i$th extremal black hole, and the charges
must all be of the same sign. It is easy to find that when $H=0$,
Eq.(31) restores to the Gibbons-Kallosh solution and when
$\alpha=0$, Eq.(31) restores to the Kastor-Traschen solution.
\section{horizons of dilaton-de Sitter spacetime}
\hspace*{7.5mm}Now let us make a discussion on the horizons of the
dilaton-de Sitter spacetime. For simplicity in mathematics, we
will concentrate on the case of the coupling constant $\alpha=1$.
To this end, make variable transformation $r=D+\sqrt{x^2+D^2}$ in
Eq.(13) and then rewrite the variable $x$ with $r$, we have
\begin{eqnarray}
d{s}^2&=&-\left(1-\frac{2M}{{D+\sqrt{r^2+D^2}}}-H^2r^2\right)
d{t}^2\nonumber\\&&+\left(1+\frac{D^2}{r^2}\right)^{-1}\left(1-\frac{2M}{{D+\sqrt{r^2+D^2}}}
-H^2r^2\right)^{-1}d{r}^2+r^2d\Omega_{2}^2.
\end{eqnarray}
It is apparent there is no singularity in the metric and when the
dilaton charge $D=0$, Eq.(32) is just the Schwarzschild-de Sitter
solution. Horizons occur whenever $g_{00}=0$. This implies that
\begin{equation}
1-\frac{2M}{{D+\sqrt{r^2+D^2}}}-H^2r^2=0.
\end{equation}
It is found that Eq.(33) gives only the cosmological horizon and
the black hole event horizon. There is no inner Cauchy horizon
present. This is different from the Reissner-Nordstr$\ddot{o}$m-de
Sitter case where three horizons are present.\\
\hspace*{7.5mm}The extremal dilaton black hole case, i.e. $M=D$,
is of particular interest. Then Eq.(33) reveals that the black
hole event horizon and the singularity disappear and only one
cosmic horizon survives
\begin{equation}
r_{cos}=\frac{1}{H}\sqrt{1-2MH}.
\end{equation}
This is also different from the extremal
Reissner-Nordstr$\ddot{o}$m-de Sitter case where two horizons are
left. Provided that
\begin{equation}
M=\frac{1}{27H}\left[9DH+8D^3H^3+\left(3+4D^2H^2\right)\sqrt{3+4D^2H^2}\right],\
\ \ \ 0< DH<\frac{1}{2},
\end{equation}
the black hole event horizon and the cosmological horizon
coalesce. That is
\begin{equation}
r_{bh,cos}=\frac{1}{3H}\sqrt{3-4D^2H^2-2DH\sqrt{3+4D^2H^2}}.
\end{equation}
Given that $DH=\frac{1}{2}$ in Eq.(36), then the only remaining
horizon $r_{bh,cos}$ also disappears. This reveals that the
presence of dilaton has important consequences on the horizons of
the black hole.
\section{conclusion and discussion}
\hspace*{7.5mm}To conclude, we have obtained the dilaton black
hole solutions in the background of (anti) de Sitter spacetime
with a new method we developed previously. The solution shows that
the cosmological constant couples the dilaton field in a
nontrivial way. The dilaton potential with respect to the
cosmological constant includes three Liouville-type potentials.
This is consistent with the discussion of Poletti and Wiltshire
that no (anti) de Sitter version of dilaton black holes exist with
only one Liouville-type potential. The resulting potentials are
also consistent with the condition
$\frac{dV}{d\phi}|_{\phi=\phi_0}=0$ which is proposed by Poletti
and Wiltshire for the existence of such solutions [18]. This type
of derived potential can be obtained when a higher dimensional
theory is compactified to four dimensions, including various super
gravity models (see[24] for a recent discussion of these aspects)
[23]. In particular, when the coupling constant
$\alpha=\pm\sqrt{\frac{1}{3}}, \pm 1, \pm\sqrt{3}$, the potential
is just the SUSY potential.\\
\hspace*{7.5mm}We also found the coordinate transformation which
recasts our solution in the Schwarzschild coordinate system. Some
discussions on the horizons in the end show that the dilaton has
important consequences on the property of the black hole. The
supersymmetry properties of the charged dilaton (anti) de Sitter
black holes as well as their relevance to the Ads/CFT
correspondence are currently under study.
 \hspace*{7.5mm}\acknowledgements This study is
supported in part by the Special Funds for Major State Basic
Research Projects and by the National Natural Science Foundation
of China. SNZ also acknowledges supports by NASA's Marshall Space
Flight Center and through NASA's Long Term Space Astrophysics
Program.

\end{document}